\documentclass[aps,preprint,tightenlines]{revtex4}%
\usepackage{amsfonts}
\usepackage{amsmath}
\usepackage{amssymb}
\usepackage{graphicx}%
\providecommand{\U}[1]{\protect \rule{.1in}{.1in}}

\begin{document}
\preprint{ }
\title{Imbalanced \emph{d}-wave superfluids in the BCS-BEC crossover regime at finite temperatures.}
\author{J.\ Tempere$^{1,\ast}$, S.N. Klimin$^{1}$, J.T. Devreese$^{1}$, V.V.
Moshchalkov$^{2}$}
\affiliation{$^{1}$Theoretische Fysica van de Vaste Stoffen (TFVS), Universiteit Antwerpen,
B-2020 Antwerpen, Belgium}
\affiliation{$^{2}$INPAC, K.U.Leuven, Celestijnenlaan 200 D, B-3001 Leuven, Belgium.}

\pacs{PACS number}

\begin{abstract}
Singlet pairing in a Fermi superfluid is frustrated when the amounts of
fermions of each pairing partner are unequal. The resulting `imbalanced
superfluid' has been realized experimentally for ultracold atomic gases with
$s$-wave interactions. Inspired by high-temperature superconductivity, we
investigate the case of $d$-wave interactions, and find marked differences
from the $s$-wave superfluid. Whereas $s$-wave imbalanced Fermi gases tend to
phase separate in real space, in a balanced condensate and an imbalanced
normal halo, we show that the $d$-wave gas can phase separate in reciprocal
space so that imbalance and superfluidity can coexist spatially. We show that
the mechanism explaining this property is the creation of polarized
excitations in the nodes of the gap. The Sarma mechanism, present only at
nonzero temperatures for the $s$-wave case, is still applicable in the
temperature zero limit for the $d$-wave case. As a result, the $d$-wave BCS
superfluid is more robust with respect to imbalance, and a region of the phase
diagram can be identified where the $s$-wave BCS superfluidity is suppressed
whereas the $d$-wave superfluidity is not. When these results are extended
into the BEC limit of strongly bound molecules, the symmetry of the order
parameter matters less. The effects of fluctuations beyond mean field is taken
into account in the calculation of the structure factor and the critical
temperature. The poles of the structure factor (corresponding to bound
molecular states) are less damped in the $d$-wave case as compared to
$s$-wave. On the BCS side of the unitarity limit, the critical temperature
$T_{c}$ follows the temperature $T^{\ast}$ corresponding to the pair binding
energy and as such will also be more robust against imbalance. Possible routes
for the experimental observation of the $d$-wave superfluidity have been discussed.

\end{abstract}
\date{\today}
\startpage{1}
\endpage{ }
\maketitle

\section{Introduction \label{sec:intro}}

In a metal exhibiting superconductivity at low temperature, the amount of
spin-up and spin-down electrons are equal, and the electron-phonon interaction
leading to Cooper pairing has a given, fixed strength. Both the population of
the spin-components and the electron-phonon interaction strength cannot be
arbitrarily tuned, and this restricts the experimental study of
superconductivity to some given values in parameter space. Nevertheless one
would like to access a much larger region of parameter space to gain insight
in pairing and the superconductivity.

Superfluid Fermi gases have recently gained a lot of interest, precisely
because of the accurate adaptability of the system parameters. The interaction
strength between the two hyperfine spin states is an adjustable parameter.
This allows to probe pairing and superfluidity in the crossover between a
Bardeen-Cooper-Schrieffer (BCS) state of weakly bound Cooper pairs and a
Bose-Einstein condensate (BEC) of tightly bound molecules
\cite{FermiBCS,FermiBEC}. Moreover, in a mixture of two hyperfine spin states
of a fermionic element, the amount of each hyperfine spin component can be
controlled experimentally. This permits to investigate the effect that a
population imbalance between the spin components has on pairing
\cite{MIT-imbal,RICE-imbal}. Not surprisingly, these recent experimental
breakthroughs \cite{MIT-imbal,RICE-imbal,FermiBCS,FermiBEC} have relaunched
the theoretical efforts to understand imbalanced Fermi superfluids in the
crossover regime \cite{theory-imbal}.

The first theoretical study of Cooper pairing in an imbalanced Fermi mixture
was performed in the context of BCS superconductors by Clogston
\cite{ClogstonPRL9}, who showed that a population imbalance destroys the
superconductivity when the imbalance in chemical potentials is of the same
order as the `balanced' order parameter. Experiments confirm that imbalance
frustrates pairing, and reveal that the excess spin component is
preferentially expelled from the superfluid \cite{FermiBCS,FermiBEC}: demixing
occurs \cite{DeSilvaPRA73}. More exotic pairing scenarios have been predicted,
most notoriously the `Fulde-Ferrell-Larkin-Ovchinnikov' scenario \cite{FFLO}
in which the Fermi spheres of the two components spontaneously deform, leading
to Cooper pairs with nonzero center-of-mass momentum.

When the temperature is raised and excitations are populated, the
superconductivity may be restored by creating a `balanced' pair condensate
with an `imbalanced' gas of excitations. This may lead to `reentrant
superconductivity' as proposed by Sarma \cite{SarmaJPCS24}. In the Sarma
state, the excess spin component is expelled from the superfluid, not in
position space, but in energy space.

In the context of the Sarma state, the case of an imbalanced Fermi gas with a
$d$-wave order parameter is particularly interesting. In the current
experiments on superfluid Fermi gases, the temperatures are low enough so that
only the $s$-wave partial wave matters, and the $d$-wave scattering is much
weaker than the $s$-wave interactions. However, the $d$-wave order parameter
has directions in momentum space where it vanishes, even at zero temperature.
This allows for a Sarma scenario where the excess spin component is expelled
from the superfluid, not in position or energy space, but in momentum space.
In this contribution, we show that $d$-wave symmetry enables the superfluid to
cope with imbalance all the way to temperature zero, using a similar scenario
as proposed by Sarma for nonzero temperature. This leads to the conclusion
that imbalance can stabilize the $d$-wave pairs with respect to the $s$-wave
pairs since the $d$-wave superfluid is more robust against population
imbalance. Moreover, since also the $d$-wave scattering length can be tuned
through the Feshbach mechanism, we investigate the $d$-wave superfluidity both
in BEC and BCS regimes.

The case of the $d$-wave pairing in the BEC/BCS crossover is also interesting
from the point of view of high-temperature superconductivity\cite{randeria},
where the order parameter is found to exhibit $d$-wave
symmetry\cite{wisenature}. The current results, derived in the context of cold
atomic gases, can also shed light on properties of the pseudogap state in the
underdoped regime. This pseudogap (which appears to have the same $d$-wave
symmetry as the order parameter\cite{WilliamsPRL87}) has been associated
either with some competing order parameter in the normal state, or with the
existence of "pre-formed pairs",\cite{alexandrovbook} where a notable
candidate is the non-condensed bipolaron\cite{alexandrov}. Also in the current
treatment, pre-formed $d$-wave pairs appear, and we present results for the
pair binding energy of these objects as a function of temperature, density and
interatomic interaction strength.

\bigskip

Our formalism of choice to treat the imbalanced Fermi gases is
path-integration. The path-integral formalism was effectively applied to study
fermionic superfluidity in dilute gases using the approach with the
Hubbard-Stratonovic transformation, choosing a saddle point, and performing
the integration over the fermionic variables. This leaves an effective action
depending on the saddle point value and the chemical potential. The effective
action can be applied to study the Fermi superfluid in optical lattices
\cite{ourlattice} or to investigate vortices in Fermi superfluids
\cite{ourvortex}. In Ref. \cite{demelo1993}, that approach was extended in
order to take into account the fluctuations around the saddle point. The
treatment starts from the partition function, which is the path integral over
fermionic (Grassmann) variables. After the introducing the auxiliary bosonic
variables and integrating over the fermionic variables, the exact expression
for the partition function from \cite{demelo1993} is the path integral over
only the boson fields with an effective bosonic action. That action is then
represented as a sum of the saddle-point contribution (which is calculated
exactly) and the contribution due to Gaussian fluctuations, which is taken
into account as a perturbation. At $T=T_{c}$, this approximation for the
fluctuations is equivalent to that of Nozi\`{e}res and Schmitt-Rink \cite{NSR}.

The further development of this idea can be found, e. g., in Refs.
\cite{Taylor2006,Randeria2007}. In Ref. \cite{Taylor2006}, the superfluid
density is derived for a uniform two-component Fermi gas in the BCS-BEC
crossover regime in the presence of an imposed superfluid flow, taking into
account pairing fluctuations in a Gaussian approximation following Ref.
\cite{NSR}. In Refs. \cite{Engelbrecht1997,Randeria2007}, the effects of
quantum fluctuations about the saddle-point solution of the BCS-BEC crossover
at $T=0$ in a dilute Fermi gas are included at the Gaussian level using the
functional integral method. In Ref. \cite{Fukushima2007}, the superfluid
density and the condensate fraction are investigated for a fermion gas in the
BCS-BEC crossover regime at finite temperatures. The fluctuation effects on
these quantities are included within a Gaussian approximation. The gas of
interacting fermions in Refs. \cite{Taylor2006,Randeria2007,Fukushima2007} is
considered for the $s$-wave pairing and with no population imbalance. A study
of the balanced $d$-wave system using the path-integral method can be found in
Ref. \cite{DM2000}. Finally, the current work applies the path-integral theory
to the \emph{imbalanced} $d$-wave superfluids, including finite-temperature
fluctuations, to show that $d$-wave pairing is particularly robust against
imbalance fluctuations.

The formalism is presented in Section \ref{sec:formalism}. In Section
\ref{sec:robustness}, we develop the mean-field approach and discuss the
resulting pair binding energy. In Section \ref{sec:Tc}, we also include the
fluctuations to treat the finite-temperature case and determine the critical
temperature for superfluidity. Near the unitarity limit where mean-field is
known to fail it is necessary to include fluctuations, but also to incorporate
the normal-state interactions in the description. The current approach
achieves this through an expansion of the action around the saddle point that
keeps terms related to the particle-hole excitations. To investigate these
excitations, we calculate in Section \ref{sec:StructureFactor} the structure
factor for the $d$-wave and compare it to the structure factor in the $s$-wave
pairing state.

\section{Formalism \label{sec:formalism}}

\subsection{$d$-wave interactions}

As in Refs. \cite{demelo1993,Randeria2007}, we start by writing down the
partition function of the interacting Fermi gas as a path integral over
Grassmann variables:
\begin{equation}
\mathcal{Z}\propto \int \mathcal{D}\bar{\psi}_{\mathbf{k},n,\sigma}%
\mathcal{D}\psi_{\mathbf{k},n,\sigma}\exp \left(  -S\right)
\end{equation}
Rather than use position and imaginary time variable, we have Grassmann fields
$\bar{\psi}_{\mathbf{k},n,\sigma}$,$\psi_{\mathbf{k},n,\sigma}$ that depend on
the wave number $\mathbf{k}$ and the fermionic Matsubara frequency $\omega
_{n}=n\pi/\beta$ with $n$ an odd integer and $\beta=1/(k_{B}T)$ the inverse
temperature. Two different hyperfine spin states are trapped so that we
include a spin quantum number $\sigma$ in the description. We'll denote the
two states as 'spin-up', $\sigma=\uparrow$ and 'spin down', $\sigma
=\downarrow$.

The action functional $S=S_{0}+S_{I}$ consists of a 'non-interacting part',
$S_{0}$, and the interaction terms, $S_{I}$. The former is given by%
\begin{equation}
S_{0}=\sum_{\mathbf{k},n}\sum_{\sigma}\left(  -i\omega_{n}+k^{2}-\mu_{\sigma
}\right)  \bar{\psi}_{\mathbf{k},n,\sigma}\psi_{\mathbf{k},n,\sigma}%
\end{equation}
where $\mu_{\sigma}$ is the chemical potential fixing the amount of atoms of
species $\sigma$, and where the summations run over all possible indices of
the Grassmann variables. We use units such that $\hbar=k_{F}=2m_{f}=1$, where
$m_{f}$ is the mass of the fermionic atoms, and $k_{F}$ is the Fermi wave
vector of the non-interacting, balanced Fermi gas with the same total number
of particles. In what follows, we will use the average chemical potential
$\mu=(\mu_{\uparrow}+\mu_{\downarrow})/2$ along with the difference in
chemical potentials $\zeta=(\mu_{\uparrow}-\mu_{\downarrow})/2$, rather than
the chemical potentials of the individual species.

The interaction terms of the action functional are written in a form that
emphasizes the \emph{pairs }of colliding fermions:%
\begin{equation}
S_{I}=\sum_{\mathbf{q},m}\sum_{\mathbf{k},n}\sum_{\mathbf{k}^{\prime
},n^{\prime}}\text{ }V_{pp}\left(  \mathbf{k,k}^{\prime}\right)  \bar{\psi
}_{\frac{\mathbf{q}}{2}+\mathbf{k},\frac{m}{2}+n,\uparrow}\bar{\psi}%
_{\frac{\mathbf{q}}{2}-\mathbf{k},\frac{m}{2}-n,\downarrow}\psi_{\frac
{\mathbf{q}}{2}-\mathbf{k}^{\prime},\frac{m}{2}-n^{\prime},\downarrow}%
\psi_{\frac{\mathbf{q}}{2}+\mathbf{k}^{\prime},\frac{m}{2}+n^{\prime}%
,\uparrow}%
\end{equation}
Here $V_{pp}$ is the interaction potential. The wave numbers in this collision
term are written as the sum of a center-of-mass wave number $\mathbf{q}$ and
the relative wave numbers $\mathbf{k,k}^{\prime}$ before and after
collision$.$ Similarly, the Matsubara frequencies are decomposed in a
center-of-mass bosonic frequency $\Omega_{m}=2m\pi/\beta$ and relative
fermionic frequencies $\omega_{n},\omega_{n^{\prime}}$. Here, we consider only
interactions that couple fermions from different hyperfine spin states. We'll
need a further assumption on the interaction potential to proceed. As in
\cite{DM2000,demelopwave} we assume that the interatomic interaction potential
can factorized as%
\begin{equation}
V_{pp}\left(  \mathbf{k,k}^{\prime}\right)  =g\Gamma \left(  \mathbf{k}\right)
\Gamma \left(  \mathbf{k}^{\prime}\right)  \label{pot}%
\end{equation}
This is possible for $s$-wave pairing%
\begin{equation}
g=g_{s},\; \Gamma_{s}\left(  \mathbf{k}\right)  =1,
\end{equation}
and also for $d$-wave pairing
\begin{equation}
g=g_{d},\; \Gamma_{d}\left(  \mathbf{k}\right)  =\frac{\left(  k/k_{1}\right)
^{2}}{\left(  1+k/k_{0}\right)  ^{5/2}}\sqrt{\frac{28\pi}{15}}Y_{2,0}\left(
\theta,\varphi \right)  .
\end{equation}
Here, $Y_{2,0}\left(  \theta,\varphi \right)  $ is the spherical harmonic, and
$k_{1},k_{0}$ are parameters fixing the range of the potential.\ The constant
$g<0$ ($g>0$) corresponds to attraction (repulsion). These constants can be
related to the $s$-and $d$-wave scattering lengths\cite{izdiasz}. The
usefulness of the factorization (\ref{pot}) lies in the fact that it allows to
rewrite the interaction terms as%
\begin{equation}
S_{I}=g\sum_{\mathbf{q},m}\bar{A}_{\mathbf{q},m}A_{\mathbf{q},m}%
\end{equation}
where we introduced the collective fields%
\begin{align}
A_{\mathbf{q},m}  &  =\sum_{n,\mathbf{k}}\frac{\Gamma \left(  \mathbf{k}%
\right)  }{\sqrt{\beta V}}\psi_{\frac{\mathbf{q}}{2}-\mathbf{k},\frac{m}%
{2}-n,\downarrow}\psi_{\frac{\mathbf{q}}{2}+\mathbf{k},\frac{m}{2}+n,\uparrow
},\nonumber \\
\bar{A}_{\mathbf{q},m}  &  =\sum_{n,\mathbf{k}}\frac{\Gamma \left(
\mathbf{k}\right)  }{\sqrt{\beta V}}\bar{\psi}_{\frac{\mathbf{q}}%
{2}+\mathbf{k},\frac{m}{2}+n,\uparrow}\bar{\psi}_{\frac{\mathbf{q}}%
{2}-\mathbf{k},\frac{m}{2}-n,\downarrow}.
\end{align}
Here $V$ is the system volume$.$ The Hubbard-Stratonovic transformation can
transform the product over these collective fields into a sum over them, at
the expense of introducing an additional functional integration:%
\begin{equation}
\mathcal{Z}\propto \int \mathcal{D}\bar{\psi}_{\mathbf{x},\tau,\sigma
}\mathcal{D}\psi_{\mathbf{x},\tau,\sigma}\int \mathcal{D}\bar{\Delta
}_{\mathbf{q},m}\mathcal{D}\Delta_{\mathbf{q},m}\exp \left(  -S\right)
\label{intermed}%
\end{equation}
with the action%
\begin{align}
S  &  =\sum_{\mathbf{k},n,\sigma}\left(  -i\omega_{n}+k^{2}-\mu_{\sigma
}\right)  \bar{\psi}_{\mathbf{k},n,\sigma}\psi_{\mathbf{k},n,\sigma}\\
&  -\sum_{m,\mathbf{q}}\left(  \frac{\bar{\Delta}_{\mathbf{q},m}%
\Delta_{\mathbf{q},m}}{g}+\bar{\Delta}_{\mathbf{q},m}A_{\mathbf{q},m}%
+\Delta_{\mathbf{q},m}\bar{A}_{\mathbf{q},m}\right)  .
\end{align}
Note that the auxiliary fields $\bar{\Delta}_{\mathbf{q},m},\Delta
_{\mathbf{q},m}$ are bosonic in nature, and characterized by the
center-of-mass pair wave number and bosonic Matsubara frequency $\Omega
_{m}=2m\pi/\beta$. The decoupling of the collective fields is necessary to
perform the functional integral over Grassmann variables, resulting in
\begin{equation}
\mathcal{Z}\propto \int \mathcal{D}\bar{\Delta}_{\mathbf{q},m}\mathcal{D}%
\Delta_{\mathbf{q},m}\exp \left(  \mathtt{tr}\ln \left[  \mathbb{G}^{-1}\left(
\mathbf{q},m;\mathbf{k},n\right)  \right]  +\frac{1}{g}\sum_{m,\mathbf{q}}%
\bar{\Delta}_{\mathbf{q},m}\Delta_{\mathbf{q},m}\right)  ,
\end{equation}
where $-\mathbb{G}^{-1}$ is the inverse Nambu tensor and the trace has to be
taken over the fermionic degrees of freedom$.$

The value where the (exponential) integrand becomes largest is called the
saddle point. Interpreting $\Delta_{\mathbf{q},m}$ as the field of bosonic
pairs, we can claim that when these pairs are condensed, the largest
contribution derives from the terms with $\Delta_{\mathbf{0},0}=\Delta$.
Performing the Bogoliubov shift, we change integration variables from
$\Delta_{\mathbf{q},m}$ to $\gamma_{\mathbf{q},m}$ where
\begin{align}
\bar{\Delta}_{\mathbf{q},m}  &  =\sqrt{V\beta}\Delta \delta_{m,0}%
\delta_{\mathbf{q},0}+\gamma_{\mathbf{q},m},\\
\bar{\Delta}_{\mathbf{q},m}  &  =\sqrt{V\beta}\bar{\Delta}\delta_{m,0}%
\delta_{\mathbf{q},0}+\bar{\gamma}_{\mathbf{q},m}.
\end{align}
If, at this point, we choose the saddle point not as the $q=0$ state, but as a
state with a finite wave number, equal to the difference between the Fermi
wave numbers of each component, we obtain the FFLO state \cite{FFLO}. Since
this has not yet been observed, we restrict the current calculations to
$\mathbf{q}=0$ pairs. This allows to split up the inverse Nambu tensor
\begin{equation}
-\mathbb{G}^{-1}\left(  \mathbf{q},m;\mathbf{k},n\right)  =\delta_{m,0}%
\delta_{\mathbf{q},0}\left[  -\mathbb{G}_{sp}^{-1}\left(  \mathbf{k},n\right)
\right]  +\mathbb{F}\left(  \mathbf{q},m;\mathbf{k}\right)  , \label{Subd}%
\end{equation}
with the saddle-point contribution is
\begin{equation}
-\mathbb{G}_{sp}^{-1}\left(  \mathbf{k},n\right)  =\left(
\begin{array}
[c]{cc}%
-i\omega_{n}+k^{2}-\mu_{\uparrow} & -\Gamma \left(  \mathbf{k}\right)  \Delta \\
-\Gamma \left(  \mathbf{k}\right)  \bar{\Delta} & -i\omega_{n}-k^{2}%
+\mu_{\downarrow}%
\end{array}
\right)  \label{Gsp}%
\end{equation}
and the fluctuation contribution is%
\begin{equation}
\mathbb{F}\left(  \mathbf{q},m;\mathbf{k}\right)  =\frac{\Gamma \left(
\mathbf{k}\right)  }{\sqrt{\beta V}}\left(
\begin{array}
[c]{cc}%
0 & -\gamma_{\mathbf{q},m}\\
-\bar{\gamma}_{-\mathbf{q},-m} & 0
\end{array}
\right)  . \label{Fluct}%
\end{equation}
We are left with the functional integration over the bosonic fields
$\gamma_{\mathbf{q},m}$ and $\bar{\gamma}_{\mathbf{q},m}$. The simplest
approximation consists in ignoring these fluctuations and setting
$\mathbb{G}=\mathbb{G}_{sp}$ -- this yields the saddle point results and will
be explored in the next subsection, \ref{sec:robustness}.B. Expanding
$\ln \left[  \mathbb{G}^{-1}\right]  $ in successive orders of $\mathbb{F}$
yields a perturbation series in $\gamma_{\mathbf{q},m}$ corresponding to an
ever increasing diagrammatic expansion, with possible Dyson resummations. The
term of order $\mathbb{F}^{2}$ is still quadratic and we calculate it in
subsection \ref{sec:robustness}.C. Up to second order:
\begin{equation}
\mathcal{Z}\propto \exp \left \{  -S_{sp}\right \}  \times \int \mathcal{D}%
\bar{\gamma}_{\mathbf{q},m}\mathcal{D}\gamma_{\mathbf{q},m}\exp \left \{
-S_{fl}\right \}  \label{1res}%
\end{equation}
with%
\begin{equation}
S_{sp}=\mathtt{tr}\ln \left[  \mathbb{G}_{sp}^{-1}\right]  -\frac{V\beta}%
{g}\bar{\Delta}\Delta
\end{equation}
and
\begin{equation}
S_{fl}=\frac{1}{2}\operatorname*{tr}\left(  \mathbb{G}_{sp}\mathbb{FG}%
_{sp}\mathbb{F}\right)  -\frac{1}{g}\sum_{\mathbf{q},m}\bar{\gamma
}_{\mathbf{q},m}\gamma_{\mathbf{q},m}.
\end{equation}
Since the partition sum is a product $\mathcal{Z}=\mathcal{Z}_{sp}%
\times \mathcal{Z}_{fl}$ , the corresponding thermodynamic potential will be a
sum of a saddle-point contribution and fluctuations: $F=F_{sp}+F_{fl}$. The
contributions are defined by $\mathcal{Z}=e^{-\beta F}$ and $\mathcal{Z}%
_{sp,fl}=e^{-\beta F_{sp,fl}}$. These thermodynamic potentials will be
necessary to calculate the two number equations in subsection
\ref{sec:robustness}.D.

\subsection{Saddle-point action}

The trace over $\mathbb{G}_{sp}^{-1}\left(  \mathbf{k},n\right)  $ can be
performed, yielding%
\[
\mathcal{Z}_{sp}=\exp \left(  -S_{sp}\right)  =\exp \left(  -\beta
F_{sp}\right)
\]
with $S_{sp}$ the saddle-point action%
\begin{equation}
S_{sp}=-\sum_{\mathbf{k},n}\ln \left[  \left(  i\omega_{n}-\zeta-E_{\mathbf{k}%
}\right)  \left(  -i\omega_{n}+\zeta-E_{\mathbf{k}}\right)  \right]
-\frac{\beta V}{g}\bar{\Delta}\Delta. \label{Ssp}%
\end{equation}
where the reader is reminded that $\zeta=(\mu_{\downarrow}-\mu_{\uparrow})/2$
is the difference in chemical potentials. The Bogoliubov energy is
\begin{equation}
E_{\mathbf{k}}=\sqrt{\left(  k^{2}-\mu \right)  ^{2}+\left \vert \Gamma \left(
\mathbf{k}\right)  \Delta \right \vert ^{2}}. \label{BE}%
\end{equation}
The sum over fermionic Matsubara frequencies can be calculated. We find for
$F_{sp}$, the saddle-point thermodynamic potential per unit volume,
\begin{equation}
\frac{F_{sp}}{V}=-\int \frac{d\mathbf{k}}{\left(  2\pi \right)  ^{3}}\left[
\frac{1}{\beta}\ln \left(  2\cosh \beta \zeta+2\cosh \beta E_{\mathbf{k}}\right)
-\xi_{\mathbf{k}}\right]  -\frac{1}{g}\left \vert \Delta \right \vert ^{2}.
\label{Fsp}%
\end{equation}
where the fermion energy is $\xi_{\mathbf{k}}=k^{2}-\mu.$ Thus, the
saddle-point result is generic for all interaction potentials of the form
(\ref{pot}).

\subsection{Quadratic fluctuations}

When the terms of order $\mathcal{O}\left(  \mathbb{F}^{3}\right)  $ and
higher are neglected in $S_{fl}$, the functional integral over $\bar{\gamma
}_{\mathbf{q},m},\gamma_{\mathbf{q},m}$ in expression (\ref{1res}) can be
performed. The result is written as%
\begin{equation}
S_{fl}=\frac{F_{fl}}{\beta V}=\frac{1}{2}\int \frac{d\mathbf{q}}{\left(
2\pi \right)  ^{3}}\sum_{m}\ln \left[  \left \vert \mathcal{M}_{1,1}\left(
\mathbf{q},i\Omega_{m}\right)  \right \vert ^{2}-\left \vert \mathcal{M}%
_{1,2}\left(  \mathbf{q},i\Omega_{m}\right)  \right \vert ^{2}\right]
\label{Ffl}%
\end{equation}
where now the trace is to be taken over bosonic Matsubara frequencies and
center of mass wave numbers. Here,
\begin{align}
\mathcal{M}_{1,1}\left(  \mathbf{q},i\Omega_{n}\right)   &  =\int
\frac{d\mathbf{k}}{\left(  2\pi \right)  ^{3}}\Gamma^{2}\left(  \mathbf{k}%
\right)  \left[  \frac{1}{2k^{2}}+\frac{\sinh \beta E_{\mathbf{k}%
-\frac{\mathbf{q}}{2}}}{2E_{\mathbf{k}-\frac{\mathbf{q}}{2}}\left(  \cosh \beta
E_{\mathbf{k}-\frac{\mathbf{q}}{2}}+\cosh \beta \zeta \right)  }\right.
\nonumber \\
&  \times \left(  \frac{\left(  i\Omega_{n}-E_{\mathbf{k}-\frac{\mathbf{q}}{2}%
}+\xi_{\mathbf{k}+\frac{\mathbf{q}}{2}}\right)  \left(  E_{\mathbf{k}%
-\frac{\mathbf{q}}{2}}+\xi_{\mathbf{k}-\frac{\mathbf{q}}{2}}\right)  }{\left(
i\Omega_{n}-E_{\mathbf{k}-\frac{\mathbf{q}}{2}}+E_{\mathbf{k}+\frac
{\mathbf{q}}{2}}\right)  \left(  i\Omega_{n}-E_{\mathbf{k}-\frac{\mathbf{q}%
}{2}}-E_{\mathbf{k}+\frac{\mathbf{q}}{2}}\right)  }\right. \nonumber \\
&  \left.  \left.  -\frac{\left(  i\Omega_{n}+E_{\mathbf{k}-\frac{\mathbf{q}%
}{2}}+\xi_{\mathbf{k}+\frac{\mathbf{q}}{2}}\right)  \left(  E_{\mathbf{k}%
-\frac{\mathbf{q}}{2}}-\xi_{\mathbf{k}-\frac{\mathbf{q}}{2}}\right)  }{\left(
i\Omega_{n}+E_{\mathbf{k}-\frac{\mathbf{q}}{2}}-E_{\mathbf{k}+\frac
{\mathbf{q}}{2}}\right)  \left(  i\Omega_{n}+E_{\mathbf{k}+\frac{\mathbf{q}%
}{2}}+E_{\mathbf{k}-\frac{\mathbf{q}}{2}}\right)  }\right)  \right]
-\lambda \left(  a\right)  , \label{M11}%
\end{align}
with the parameters $\lambda \left(  a\right)  $ which describe the coupling
strength for the $s$-wave and $d$-wave pairings\cite{izdiasz}:%
\begin{equation}
\lambda_{s}\left(  a_{s}\right)  =\frac{1}{8\pi a_{s}},\; \lambda_{d}\left(
a_{d}\right)  =\frac{2}{\pi a_{d}^{5}}, \label{lambda}%
\end{equation}
and%
\begin{align}
\mathcal{M}_{1,2}\left(  \mathbf{q},i\Omega_{n}\right)   &  =-\left \vert
\Delta \right \vert ^{2}\int \frac{d\mathbf{q}}{\left(  2\pi \right)  ^{3}}%
\Gamma^{2}\left(  \mathbf{k}\right)  \Gamma \left(  \mathbf{k}+\frac
{\mathbf{q}}{2}\right)  \Gamma \left(  \mathbf{k}-\frac{\mathbf{q}}{2}\right)
\nonumber \\
&  \times \frac{\sinh \beta E_{\mathbf{k}-\frac{\mathbf{q}}{2}}}{2E_{\mathbf{k}%
-\frac{\mathbf{q}}{2}}\left(  \cosh \beta E_{\mathbf{k}-\frac{\mathbf{q}}{2}%
}+\cosh \beta \zeta \right)  }\nonumber \\
&  \times \left(  \frac{1}{\left(  i\Omega_{n}-E_{\mathbf{k}-\frac{\mathbf{q}%
}{2}}+E_{\mathbf{k}+\frac{\mathbf{q}}{2}}\right)  \left(  i\Omega
_{n}-E_{\mathbf{k}-\frac{\mathbf{q}}{2}}-E_{\mathbf{k}+\frac{\mathbf{q}}{2}%
}\right)  }\right. \nonumber \\
&  \left.  +\frac{1}{\left(  i\Omega_{n}+E_{\mathbf{k}-\frac{\mathbf{q}}{2}%
}-E_{\mathbf{k}+\frac{\mathbf{q}}{2}}\right)  \left(  i\Omega_{n}%
+E_{\mathbf{k}-\frac{\mathbf{q}}{2}}+E_{\mathbf{k}+\frac{\mathbf{q}}{2}%
}\right)  }\right)  . \label{M22}%
\end{align}

In the particular case of the $s$-wave pairing and of the balanced fermion
gas, (\ref{M11}) and (\ref{M22}) are equivalent to the matrix elements derived
in Ref. \cite{Engelbrecht1997}. In the present treatment, as distinct from
Refs. \cite{Engelbrecht1997,Randeria2007}, we do not assume the
low-temperature limit.

\subsection{Gap and number equations}

The \textbf{gap equation} is determined by $S_{sp}$ alone, through $\delta
S_{sp}/\delta \Delta=0$. The gap equation can be written in a unified form for
the $s$-wave and $d$-wave pairings,%
\begin{equation}
\int \frac{d\mathbf{k}}{\left(  2\pi \right)  ^{3}}\left \vert \Gamma \left(
\mathbf{k}\right)  \right \vert ^{2}\left(  \frac{\sinh \beta E_{\mathbf{k}}%
}{2E_{\mathbf{k}}\left(  \cosh \beta E_{\mathbf{k}}+\cosh \beta \zeta \right)
}-\frac{1}{2k^{2}}\right)  +\lambda \left(  a\right)  =0. \label{gap2}%
\end{equation}
The \textbf{number equations} are determined from the thermodynamic potential
through%
\begin{align}
\left(  \frac{\partial F}{\partial \mu}\right)  _{T,V,\Delta}  &
=-n,\label{dens}\\
\left(  \frac{\partial F}{\partial \zeta}\right)  _{T,V,\Delta}  &  =-\delta n,
\label{diff}%
\end{align}
where $n=n_{\uparrow}+n_{\uparrow}$ is the total local density, and $\delta
n=n_{\uparrow}-n_{\uparrow}$ is the local population imbalance. For a finite
temperature below $T_{c}$, the chemical potentials $\mu$ and $\zeta$, and the
gap $\Delta$ are determined self-consistently as a solution of the gap
equation (\ref{gap2}) coupled with the number equations (\ref{dens}) and
(\ref{diff}). In principle, we can write the exact thermodynamic potential
$F=F_{sp}+F_{fl}+F_{other}$ where $F_{sp}$ and $F_{fl}$ are given by
expressions (\ref{Fsp}) and (\ref{Ffl}), and $F_{other}$ comes from the
contributions of all higher order terms, $\mathcal{O}\left(  \mathbb{F}%
^{3}\right)  ,$ in the exact action. The local density and the local
population imbalance can be written as a sum of several contributions,%
\begin{align}
n  &  =n_{sp}+n_{fl}+n_{other},\\
\delta n  &  =\delta n_{sp}+\delta n_{fl}+\delta n_{other},
\end{align}
where $n_{sp}$ and $\delta n_{sp}$ are the saddle-point results, $n_{fl}$ and
$\delta n_{fl}$ are the fluctuation contributions, and $n_{other}$, $\delta
n_{other}$ are higher-order fluctuation contributions to the density and
population imbalance, which are neglected in the present treatment.

The saddle-point contributions to the density and to the population imbalance
are obtained using the saddle-point term of the thermodynamic potential
(\ref{Fsp}) and Eqs. (\ref{dens}) and (\ref{diff}):%
\begin{align}
n_{sp}  &  =\frac{1}{2\pi^{2}}\int_{0}^{\infty}k^{2}dk\left(  1-\frac
{\varepsilon_{\mathbf{k}}}{E_{\mathbf{k}}}\frac{\sinh \left(  \beta
E_{\mathbf{k}}\right)  }{\cosh \left(  \beta \zeta \right)  +\cosh \left(  \beta
E_{\mathbf{k}}\right)  }\right)  ,\label{deq1}\\
\delta n_{sp}  &  =\frac{1}{2\pi^{2}}\int_{0}^{\infty}k^{2}dk\frac
{\sinh \left(  \beta \zeta \right)  }{\cosh \left(  \beta \zeta \right)
+\cosh \left(  \beta E_{\mathbf{k}}\right)  }. \label{deq2}%
\end{align}

The fluctuation contribution to the number equations is determined on the
basis of the fluctuation contribution to the thermodynamic potential:%
\begin{align}
n_{fl}  &  =-\frac{1}{\beta}\int \frac{d\mathbf{q}}{\left(  2\pi \right)  ^{3}%
}\sum_{n=-\infty}^{\infty}J\left(  \mathbf{q},i\Omega_{n}\right)
,\label{nfl1}\\
\delta n_{fl}  &  =-\frac{1}{\beta}\int \frac{d\mathbf{q}}{\left(  2\pi \right)
^{3}}\sum_{n=-\infty}^{\infty}K\left(  \mathbf{q},i\Omega_{n}\right)  ,
\label{dnfl1}%
\end{align}
where the functions $J\left(  \mathbf{q},z\right)  $ and $K\left(
\mathbf{q},z\right)  $ for a complex argument $z$ are given by
\begin{align}
J\left(  \mathbf{q},z\right)   &  =\frac{1}{\Gamma \left(  \mathbf{q},z\right)
}\left[  \mathcal{M}_{1,1}\left(  \mathbf{q},-z\right)  \frac{\partial
\mathcal{M}_{1,1}\left(  \mathbf{q},z\right)  }{\partial \mu}-\mathcal{M}%
_{1,2}\left(  \mathbf{q},-z\right)  \frac{\partial \mathcal{M}_{1,2}\left(
\mathbf{q},z\right)  }{\partial \mu}\right]  ,\label{FJ}\\
K\left(  \mathbf{q},z\right)   &  =\frac{1}{\Gamma \left(  \mathbf{q},z\right)
}\left[  \mathcal{M}_{1,1}\left(  \mathbf{q},-z\right)  \frac{\partial
\mathcal{M}_{1,1}\left(  \mathbf{q},z\right)  }{\partial \zeta}-\mathcal{M}%
_{1,2}\left(  \mathbf{q},-z\right)  \frac{\partial \mathcal{M}_{1,2}\left(
\mathbf{q},z\right)  }{\partial \zeta}\right]  , \label{FK}%
\end{align}
with%
\begin{equation}
\Gamma \left(  \mathbf{q},z\right)  =\mathcal{M}_{1,1}\left(  \mathbf{q}%
,z\right)  \mathcal{M}_{1,1}\left(  \mathbf{q},-z\right)  -\mathcal{M}%
_{1,2}\left(  \mathbf{q},z\right)  \mathcal{M}_{1,2}\left(  \mathbf{q}%
,-z\right)  . \label{Gamma}%
\end{equation}
The functions $\mathcal{M}_{1,1}\left(  \mathbf{q},z\right)  $ and
$\mathcal{M}_{1,2}\left(  \mathbf{q},z\right)  $ of the complex argument $z$
are analytical in the complex $z$-plane except the branching line, which lies
at the real axis $z=\omega$. Similarly to Refs. \cite{demelo1993,Nozieres1985}%
, the summations over the boson Matsubara frequencies in (\ref{nfl1}) and
(\ref{dnfl1}) are converted to the contour integrals in the complex plane as
described in the Appendix. Here, we write down the final result for the
fluctuation contributions to $n$ and $\delta n$:%
\begin{align}
n_{fl}  &  =-\int \frac{d\mathbf{q}}{\left(  2\pi \right)  ^{3}}\left(  \frac
{1}{\pi}\int_{-\infty}^{\infty}\operatorname{Im}\left[  \frac{J\left(
\mathbf{q},\omega+i\gamma \right)  }{e^{\beta \left(  \omega+i\gamma \right)
}-1}\right]  d\omega+\frac{1}{\beta}\sum_{n=-n_{0}}^{n_{0}}J\left(
\mathbf{q},i\Omega_{n}\right)  \right)  ,\label{nfl}\\
\delta n_{fl}  &  =-\int \frac{d\mathbf{q}}{\left(  2\pi \right)  ^{3}}\left(
\frac{1}{\pi}\int_{-\infty}^{\infty}\operatorname{Im}\left[  \frac{K\left(
\mathbf{q},\omega+i\gamma \right)  }{e^{\beta \left(  \omega+i\gamma \right)
}-1}\right]  d\omega+\frac{1}{\beta}\sum_{n=-n_{0}}^{n_{0}}K\left(
\mathbf{q},i\Omega_{n}\right)  \right)  . \label{dnfl}%
\end{align}
Here, the number $n_{0}$ is chosen arbitrarily, and the parameter $\gamma$
lies in the range $\Omega_{n_{0}}<\gamma<\Omega_{n_{0}+1}$.

In particular, if one chooses $n_{0}=0,$ the formula (\ref{sum}) leads to the
expression for the fluctuation contribution to the fermion density similar to
that derived in Ref. \cite{demelo1993}:%
\begin{equation}
n_{fl}=\frac{1}{\pi}\int \frac{d\mathbf{q}}{\left(  2\pi \right)  ^{3}q^{2}}%
\int_{-\infty}^{\infty}d\omega S\left(  \mathbf{q},\omega \right)  .
\label{n0fl}%
\end{equation}
Here, the structure factor is%
\begin{equation}
S\left(  \mathbf{q},\omega \right)  =-\frac{q^{2}\operatorname{Im}\left[
J\left(  \mathbf{q},\omega+i\delta \right)  \right]  }{e^{\beta \omega}-1},\;
\delta \rightarrow+0. \label{SF}%
\end{equation}
The results obtained in the present section extends the path-integral approach
of Ref. \cite{Randeria2007} to the case of the $d$-wave pairing and of an
imbalanced Fermi gas at arbitrary temperatures. In agreement with the proof
made in Ref. \cite{Nozieres1985}, the function%
\begin{equation}
Q\left(  \mathbf{q},\omega \right)  \equiv \lim_{\delta \rightarrow+0}\left \{
\operatorname{Im}\left[  J\left(  \mathbf{q},\omega+i\delta \right)  \right]
\right \}
\end{equation}
at $T=T_{c}$ is equal to zero at $\omega=0$. Furthermore, $Q\left(
\mathbf{q},\omega \right)  $ changes its sign as $\omega$ passes through
$\omega=0$. This is necessary to ensure that the relative contribution to the
fluctuation density from excitations with given $\left(  \mathbf{q}%
,\omega \right)  $ remains positive; this contribution is proportional to
$S\left(  \mathbf{q},\omega \right)  $.

\section{Robustness of the $d$-wave pair binding energy \label{sec:robustness}%
}

\begin{figure}
[ptbh]
\begin{center}
\includegraphics[
height=3.0027in,
width=3.1648in
]%
{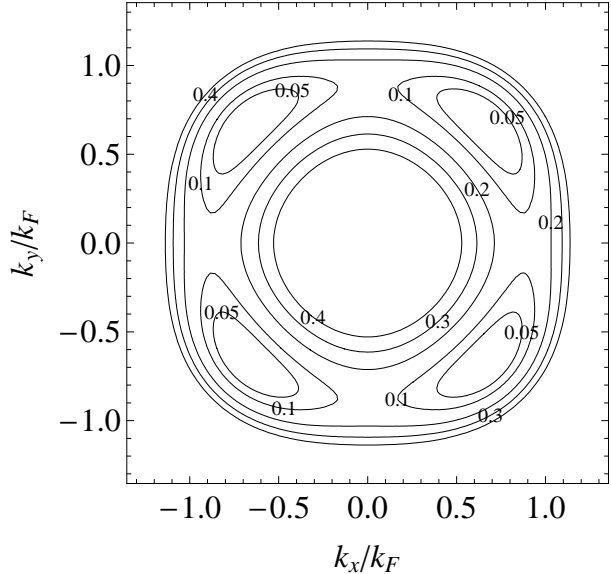}%
\caption{Even a small difference between the chemical potential leads to
population imbalance. This plot shows the contours of $E_{\mathbf{k}}=\zeta$
for different values of $\zeta$ in the $k_{x},k_{y}$ plane, for a $d$-wave
interaction characterized by $k_{0}=k_{1}=10k_{F}$. Within the regions
$E_{\mathbf{k}}<\zeta$, spin-polarized Bogoliubov excitations are present that
carry the excess spin component of the imbalanced gas. }%
\label{fig1}%
\end{center}
\end{figure}

First we look at the saddle-point results for temperature zero, in order to
investigate the pair binding energy. In the limit of temperature zero
$\beta \rightarrow \infty$, the gap equation becomes%
\begin{equation}
-\frac{2}{\pi(k_{F}a_{d})^{5}}=\int \frac{d\mathbf{k}}{(2\pi)^{3}}\left \{
\frac{\Theta(E_{\mathbf{k}}>\zeta)}{2E_{\mathbf{k}}}-\frac{1}{2k^{2}}\right \}
|\Gamma(\mathbf{k})|^{2} \label{spgap}%
\end{equation}
with $\Theta$ the logical Heaviside function. Simultaneously the two
saddle-point number equations (\ref{deq1}),(\ref{deq2}) become
\begin{align}
\frac{1}{3\pi^{2}}  &  =\int \frac{d\mathbf{k}}{(2\pi)^{3}}\left \{
1-\Theta(E_{\mathbf{k}}>\zeta)\frac{\varepsilon_{\mathbf{k}}}{E_{\mathbf{k}}%
}\right \} \label{spnum1}\\
\frac{1}{3\pi^{2}}\frac{\delta n}{n}  &  =\int \frac{d\mathbf{k}}{(2\pi)^{3}%
}\Theta(E_{\mathbf{k}}<\zeta). \label{spnum2}%
\end{align}
The $\Theta(E_{\mathbf{k}}<\zeta)$ function cuts off all the wave numbers with
energy less than $\zeta$. These are shown in Fig. \ref{fig1}. Near
$\{k_{x},k_{y}\}=\{ \sqrt{2},\sqrt{2}\}k_{F}$ the gap vanishes, and
excitations are always present. To have non-zero imbalance in an $s$-wave
superfluid, $\zeta$ has to be of the order of $\left \vert \Delta \right \vert $.
This is the Clogston limit, and superfluidity will break down. However, for
the $d$-wave superfluid all values of $\zeta$ lead to imbalance, and small
values of $\zeta$ do not destroy superfluidity.%

Solving for a given imbalance and interaction strength $a_{d}$ the
saddle-point number and gap equations (\ref{spgap})-(\ref{spnum2}), we can
derive the saddle-point value $\Delta$. This value is necessary to compute the
fluctuation effects, but it has an interpretation by itself, namely as the
pair binding energy. A corresponding temperature $T^{\ast}=\left \vert
\Delta \right \vert /k_{B}$ can be associated with the pair binding energy. In
the BCS\ limit, superfluidity is destroyed by breaking up Cooper pairs. Thus,
the transition temperature is determined by the binding energy of the Cooper
pairs and $T_{c}\approx T^{\ast}$. However, in the BEC limit, superfluidity is
destroyed not by breaking up the bosonic molecules, but by phase fluctuations,
and typically $T_{c}\ll T^{\ast}$. The BEC limit, with its tightly bound
molecules, is relatively insensitive to the addition of atoms of one of the
spin species: the imbalanced system can be described as a mixture of fermionic
atoms and bosonic molecules. The BCS limit, however, is very sensitive to
imbalance. Since in the BCS limit $T_{c}$ is directly related to the pair
binding energy $\left \vert \Delta \right \vert $, we gain insight in the effect
of imbalance on $s$- and $d$-wave superfluids through the saddle-point gap.%

\begin{figure}
[ptb]
\begin{center}
\includegraphics[
height=4.5722in,
width=3.0258in
]%
{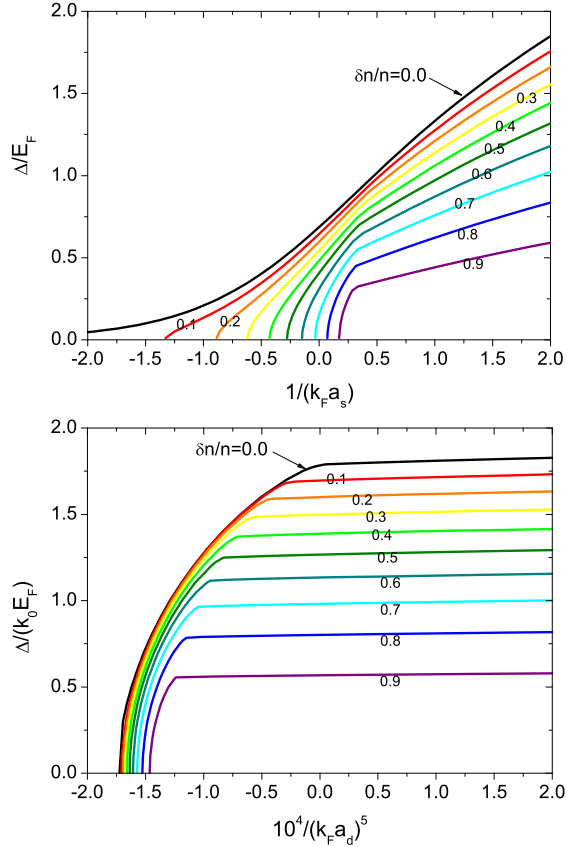}%
\caption{(color online) The dependence of the pair binding energy on the
interaction strength for the $s$-wave (top) and $d$-wave (bottom) interaction
is influenced by the imbalance $\delta n/n=0.0$ ... $0.9$. On the BCS side,
the imbalance destroys $s$-wave Cooper pairs for all values of $a_{s}<0$, but
fails to break up the $d$-wave Cooper pairs.}%
\label{fig2}%
\end{center}
\end{figure}

The result for $\left \vert \Delta \right \vert $ is shown in Fig. \ref{fig2} for
$s$-wave (top panel) and $d$-wave (bottom panel) pairing. There are some
notable differences between $s$- and $d$-wave results. Firstly, for the
$d$-wave interaction the $x$-axis is a function of $1/(k_{F}a_{d})^{5}$ in
stead of \thinspace$1/(k_{F}a_{s})$. This means that the $d$-wave scattering
length should be much closer to resonance as compared to the $s$-wave case
before superfluidity enters the resonant regime. The absolute scale still
depends on $k_{0}$, related to the range of the interaction potential. Also
the scale of the y-axis in the graph (representing $\left \vert \Delta
\right \vert )$ has this dependence on the details of the potential embodied in
$\Gamma(\mathbf{k})$.

A second difference between $s$- and $d$-wave resonant pairing, is that for
$s$-wave interactions we find that there is pairing for all values of
$a_{s}<0$. For $d$-wave interactions, it is no longer true that for any
attractive potential there is pairing. There needs to be a fatal attraction
before pairing occurs on the BCS side. The BEC side, however, is more or less
the same for $s$- and $d$-wave. Deep in the BEC regime it indeed should not
matter whether we have $s$-wave or $d$-wave internal parameter.

A third difference, is that the $d$-wave order parameter on the BCS side is
much more robust to imbalance than the $s$-wave order parameter. For all
negative scattering lengths, there exists a critical imbalance that destroys
superfluidity in the $s$-wave system. However, in the $d$-wave case, there is
a range of negative scattering lengths for which the pairs remain bound up to
the maximal imbalance. This confirms our intuition that the excess spin
component can be nicely stowed away in the minima of the gap, near the
$k/k_{F}=\{1/\sqrt{2},1/\sqrt{2}\}$ point. At these points, the gap vanishes
naturally and it does not cost any energy to make excitations or to store
broken pairs.

The $d$-wave order parameter shows similarity to the $s$-wave when some
imbalance already present (the $\delta n/n=0$ curve looks like the $s$-wave
curve for nonzero imbalance), but it is much more robust to imbalance. One can
imagine increasing imbalance in such a way that it suppresses the $s$-wave
pairing channel and still allows the $d$-wave pairing channel.

\section{Critical temperature for the $d$-wave pairing in the region of the
BCS-BEC crossover \label{sec:Tc}}

\begin{figure}
[ptbh]
\begin{center}
\includegraphics[
height=5.5884in,
width=5.9681in
]%
{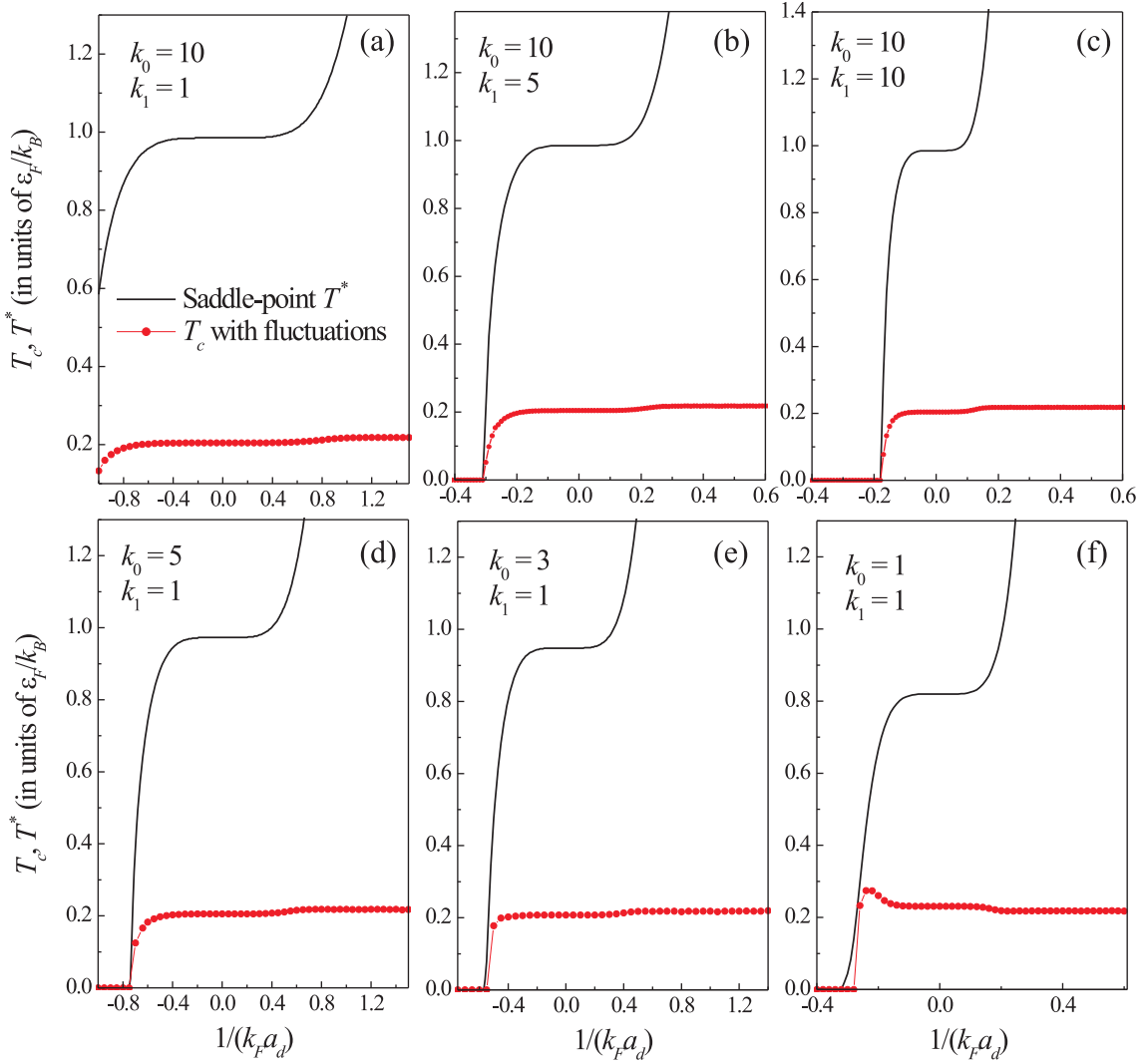}%
\caption{(color online) Critical temperature for the fermion system with the
$d$-wave scattering calculated taking into account the fluctuations around the
saddle point as a function of the inverse scattering length (red full dots).
The saddle-point critical temperature $T^{\ast}$ is plotted with the solid
black curves.}%
\label{fig:Tc}%
\end{center}
\end{figure}

At finite temperatures, both phase fluctuations and amplitude fluctuations in
$\Delta$ are important. The amplitude fluctuations dominate the thermodynamics
in the BCS regime, whereas the phase fluctuations dominate in the BEC regime.
This will be borne out in more detail by a study of the structure factor, in
Sec. IV. For a given temperature $T$, density $n$ and density imbalance
$\delta n$, we can solve the gap and number equations numerically and
determine $\Delta,\mu,\zeta$. The critical temperature can be found as the
temperature where $\Delta$ vanishes. In Fig. \ref{fig:Tc}, we plot the
critical temperature in the case of the $d$-wave scattering as a function of
the inverse scattering length $1/(k_{F}a_{d})$. The saddle-point results for
the pair breaking temperature $T^{\ast}=\left \vert \Delta \right \vert /k_{B}$
are plotted with the solid black curves, and the values of $T_{c}$ calculated
taking into account the fluctuations are plotted with red full dots.%

For all considered values of the parameters $k_{0},k_{1}$ of the $d$-wave
scattering potential, we can see three following regions of $1/(k_{F}a_{d})$,
with different behavior of $T_{c}$.\newline(1) A region corresponding to the
weak-coupling regime (at $1/(k_{F}a_{d})<0$). In this regime, with increasing
$1/(k_{F}a_{d})$, the critical $T_{c}$ starts from the value $T_{c}=0$ at a
certain value $1/(k_{F}a_{d})$, and rapidly increases. This is consistent with
the finding in the previous section, that a critical strength of the
interatomic interaction is required before pair formation occurs. \newline(2)
The region of the \textquotedblleft plateau\textquotedblright \ around the
unitarity point $1/(k_{F}a_{d})=0$, where $T_{c}$ varies extremely
slowly.\newline(3) The region corresponding to the strong-coupling regime (at
$1/(k_{F}a_{d})>0$), where $T_{c}$ tends to the finite value $T_{c}%
\approx0.218.$ This is the same value as obtained in Ref. \cite{demelo1993}
for the $s$-wave case. Indeed we expect in the deep BEC\ limit the details of
the internal structure of the molecule to be of secondary importance.

Compared to the case of the $s$-wave scattering \cite{demelo1993}, the
dependence $T_{c}\left(  a_{d}\right)  $ for the $d$-wave scattering has a
broad plateau around the point $1/(k_{F}a_{d})=0$ both for the saddle-point
results and for those taking into account the fluctuations. This plateau is
explained by the fact that the factor $1/(k_{F}a_{d})$ enters the gap and
number equations through its fifth power, $1/(k_{F}a_{d})^{5}$, which varies
very slowly as compared to the case of the $s$-wave scattering in the
unitarity region. Another difference with the $s$-wave case is that there is a
critical value of $a_{d}$ so that both $T^{\ast}$ and $T_{c}$ turn to zero.
This means a minimum strength of the attraction is necessary to achieve
pairing in the $d$-wave, whereas in the $s$-wave case for all values of the
(negative) scattering length one has pairing.

As the BEC limit is approached, $T^{\ast}$ strongly increases, while $T_{c}$
tends to a constant value. Again the behavior in the BEC limit is similar to
that of the $s$-wave case, as can be expected. In the BCS regime,
$T_{c}\approx T^{\ast}$ as anticipated in the previous subsection. For
$k_{0}=10$, 5 and 3, $T_{c}$ is a slightly increasing function of
$1/(k_{F}a_{d})$ at positive $1/k_{F}a_{d}$ and remains everywhere lower than
$T^{\ast}.$ For $k_{0}=1$, however, we see that $T_{c}$ achieves a maximum at
a negative value of $1/(k_{F}a_{d})$ and then decreases to the BEC limit. This
behavior of $T_{c}$ shows that at $k_{0}=10$, 5 and 3, the anti-crossing of
BCS and BEC regimes occurs at $a_{d}>0$, and that with decreasing $k_{0}$, the
region of the anti-crossing of BCS and BEC regimes shifts to lower values of
the inverse scattering length.

\section{Structure factor for the $s$-wave and $d$-wave pairings
\label{sec:StructureFactor}}

In the case with $\gamma \rightarrow+0,$ and for a balanced Fermi gas, the
contribution $n_{fl}$ given by (\ref{nfl}) is expressed through the integral
(\ref{n0fl}) with the structure factor $S\left(  \mathbf{q},\omega \right)  $
given by (\ref{SF}). The structure factor is of particular interest, because
it represents the spectrum of the elementary excitations of the fermion gas
below $T_{c}$. Further on, we analyze the structure factor $S\left(
\mathbf{q},\omega \right)  $ at $T=T_{c}$ and the excitation spectra for the
cases of the \emph{s}-wave and \emph{d}-wave scattering, comparing to existing
results for the $s$-wave case \cite{OhashiPRA67}. Whereas the poles of the
single-particle Green's function can be associated with single-particle
excitations, the poles of $S\left(  \mathbf{q},\omega \right)  $ are related in
the present case to the two-particle bound state. Note that the next term in
the fluctuation expansion, proportional to the fourth power of $\Delta$, gives
rise to a spectral function the poles of which are related to the collective
excitations of these bound modes.%

\begin{figure}
[h]
\begin{center}
\includegraphics[
height=4.3336in,
width=5.5227in
]%
{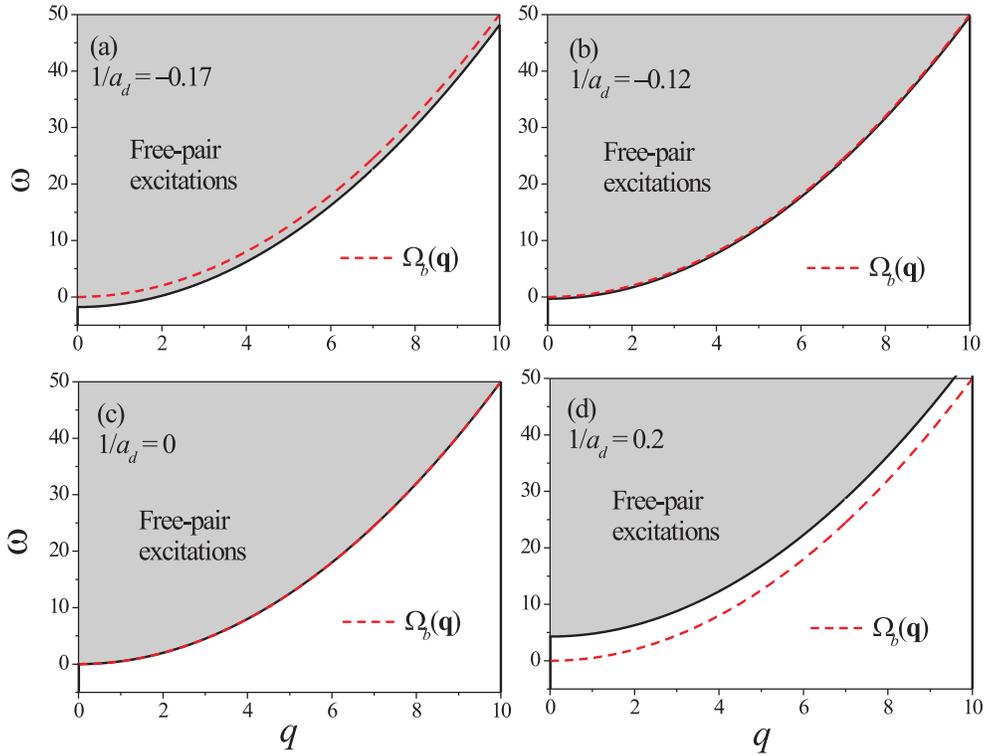}%
\caption{(color online) Excitation region of a gas of interacting fermions in
$\left(  q,\omega \right)  $-space (the case of the \emph{d}-wave scattering)
at $T=T_{c}$ for $k_{0}=10$, $k_{1}=10$, and $\cos \theta=1/2$. The shaded area
shows the continuum of the two-particle excitations. The black curve denotes
the lower-frequency bound of the damping area. The dashed red curve shows the
points given by $\omega=\Omega_{b}\left(  \mathbf{q}\right)  $, where
$\Omega_{b}\left(  \mathbf{q}\right)  =\omega_{b}\left(  \mathbf{q}\right)
-2\mu$ with the energy $\omega_{b}\left(  \mathbf{q}\right)  $ of the two-body
bound state.}%
\label{excitations}%
\end{center}
\end{figure}

In Fig. \ref{excitations}, we have plotted the excitation region of the
fermion gas in $\left(  q,\omega \right)  $-space for the \emph{d}-wave
scattering at different values of the inverse scattering length $1/a_{d}$ and
at $T=T_{c}$. The solid black curve denotes the the lower bound $\omega
_{0}\left(  q\right)  =q^{2}/2-2\mu$ for the continuum of free two-particle
excitations, which is determined by the inequality $\omega>\omega_{0}\left(
q\right)  $. The dashed red curve corresponds to the solution of the equation
$\omega=\Omega_{b}\left(  \mathbf{q}\right)  $, where $\Omega_{b}\left(
\mathbf{q}\right)  $ is the pole of the structure factor $S\left(
\mathbf{q},\omega \right)  $. In the case when $\Omega_{b}\left(
\mathbf{q}\right)  <\omega_{0}\left(  q\right)  $, i. e., when the pole
$\Omega_{b}\left(  \mathbf{q}\right)  $ lies outside the continuum of
two-particle excitations, $\Omega_{b}\left(  \mathbf{q}\right)  $ is given by
$\Omega_{b}\left(  \mathbf{q}\right)  =\omega_{b}\left(  \mathbf{q}\right)
-2\mu,$ where $\omega_{b}\left(  \mathbf{q}\right)  $ is the energy of the
two-body bound state \cite{demelo1993}. In the strong-coupling limit, the
energy of the two-body bound state tends to $\omega_{b}\left(  \mathbf{q}%
\right)  =-E_{b}+q^{2}/2$, where $E_{b}$ is the pair binding energy, which in
this limit and at $T=T_{c}$ tends to $\left(  -2\mu \right)  $. For a
sufficiently weak coupling, the pole $\Omega_{b}\left(  \mathbf{q}\right)  $
lies within the continuum, and therefore the two-body bound state is damped.

In the case of $d$-wave scattering, for all considered values of $q$, the
two-body bound states are damped at $1/a_{d}<0$ [Fig. \ref{excitations} (a,
b)] and non-damped at $1/a_{d}\gtrapprox0$ [Fig. \ref{excitations} (d)]. For
$1/a_{d}=0$ [Fig. \ref{excitations} (c)], $\omega_{0}\left(  q\right)  $ and
$\Omega_{b}\left(  q\right)  $ practically coincide. This allows to interpret,
in the case of $d$-wave scattering, the value $1/a_{d}=0$ as the boundary
between the regimes of the BCS-pairing (for $1/a_{d}<0$) and the BEC-pairing
(for $1/a_{d}\gtrapprox0$).

\begin{figure}
[h]
\begin{center}
\includegraphics[
height=2.3263in,
width=5.0868in
]%
{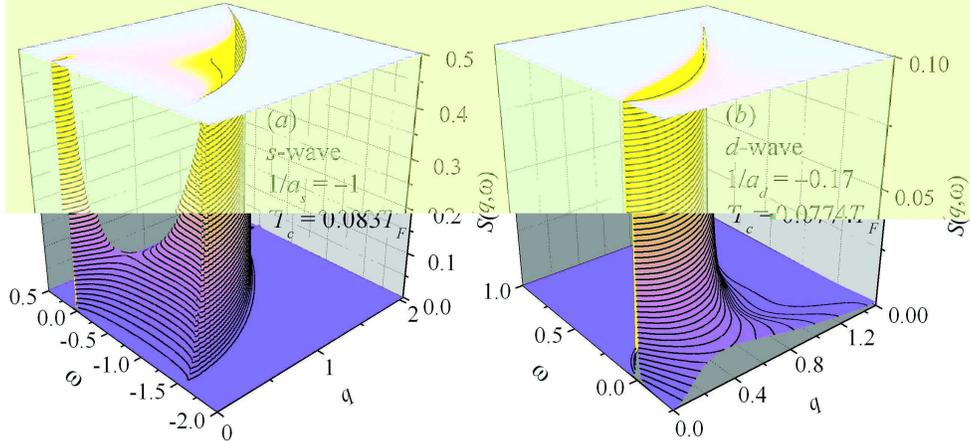}%
\caption{(color online) 3D plot of the structure factor $S\left(
q,\omega \right)  $ for the $s$-wave scattering (\emph{a}) and for the $d$-wave
scattering (\emph{b}) at $T=T_{c}$ in the weak-coupling regime. The critical
temperatures are given in units of the Fermi temperature $T_{F}\equiv \hbar
^{2}k_{F}^{2}/\left(  2mk_{B}\right)  $. At the top, there are the contour
density plots for $S\left(  q,\omega \right)  $.}%
\label{SFweak}%
\end{center}
\end{figure}

The 3D plots in Fig. \ref{SFweak} represent the structure factors for the
\emph{s}-wave and \emph{d}-wave pairing at weak-coupling. Because the
scattering potential for the $d$-wave scattering is angle-dependent, the
structure factor $S\left(  \mathbf{q},\omega \right)  $ depends on three
variables: $S\left(  \mathbf{q},\omega \right)  =S\left(  q,\cos \theta
,\omega \right)  $. Here, we discuss the results for $S\left(  q,\omega \right)
$ averaged over the directions,%
\begin{equation}
S\left(  q,\omega \right)  \equiv \frac{1}{2}\int_{0}^{\pi}S\left(
\mathbf{q},\omega \right)  \sin \theta d\theta.
\end{equation}
%

In Fig. \ref{SFweak}~(\emph{a}), the structure factor for the case of $s$-wave
pairing is shown for $1/a_{s}=-1$, which lies on the BCS side of the
resonance. In the BCS regime, the poles corresponding to the two-body pair
excitations are damped since they lie in the continuum area. The spectral
weight of those poles in the overall structure factor is significant only at
small wave vectors. This results in a peak around $q=0$ in Fig. \ref{SFweak}%
~(\emph{a}). Furthermore, there is a distinctive extremum of the structure
factor at the boundary of the continuum of two-particle excitations. In Fig.
\ref{SFweak}~(\emph{b}) we switch from $s$-wave to $d$-wave pairing, but
remain within on BCS side of the resonance: this panel shows the structure
factor for the $d$-wave pairing is for $1/a_{d}=-0.17$. This value of the
$d$-wave scattering length is close to the lowest value of the inverse
scattering length at which pairing can occur. Also in the $d$-wave BCS regime,
the two-body bound states are damped and therefore the peak corresponding to
the two-body bound excitations has a finite width. However, the width of that
peak in the case of the $d$-wave scattering is relatively low. From this we
can see that in the case of the $d$-wave scattering, the two-body bound state
plays a significant role even in the weak-coupling regime. For the $d$-wave
scattering, as distinct from the $s$-wave scattering, the BCS pairing
mechanism can be realized only in a narrow range of the inverse scattering
length close to the lowest value of $1/a_{d}$ from those, for which pairing
can occur.%

\begin{figure}
[h]
\begin{center}
\includegraphics[
height=2.3376in,
width=5.0756in
]%
{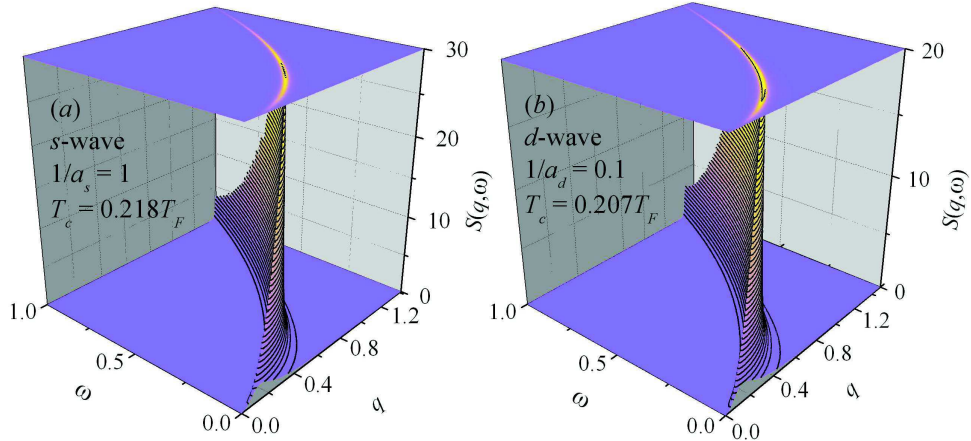}%
\caption{(color online) 3D plot of the structure factor $S\left(
q,\omega \right)  $ for the $s$-wave scattering (\emph{a}) and for the $d$-wave
scattering (\emph{b}) at $T=T_{c}$ in the strong-coupling regime.}%
\label{SFstrong}%
\end{center}
\end{figure}

Fig. \ref{SFstrong} describes the strong-coupling case (on the BEC side of the
resonance) where there is a non-damped isolated pole in the structure factor.
Therefore, the structure factor in the strong-coupling regime contains a
$\delta$-like peak, which lies outside the continuum of free-pair excitations.
In order to visualize those $\delta$-like peaks, we use a finite damping
parameter $\gamma=0.01$. In the strong-coupling regime, the regular part of
$S\left(  \mathbf{q},\omega \right)  $ is negligibly small with respect to the
main contribution due to the aforesaid isolated pole, which describes the BEC pairing.

\section{Discussion and conclusions \label{conclusions}}

\subsection{The unitarity limit}

A first point to discuss, is the predicament of mean-field theory in the
unitarity limit, $1/a\rightarrow0$. It is crucial that fluctuations are taken
into account, as we have done in the previous section. Higher-order
fluctuation contributions can be taken into account, and many different
approaches were developed for the balanced $s$-wave Fermi superfluid. These
approaches differ in the types of higher-order processes they take into
account. Strinati and co-workers \cite{strinati} work diagrammatically to
improve on the results obtained by Nozi\`{e}res and Schmitt-Rink for $T=T_{c}%
$. Levin and co-workers \cite{levin} construct a finite temperature theory
similar to that of Strinati and co-workers but include different diagrams in
the summation. Alternatively, quantum Monte Carlo simulations \cite{MC} can be
used to obtain results on the crossover physics the balanced $s$-wave Fermi
superfluid and put the crossover theories to the test . At the unitarity limit
the existing theories do not succeed to find the free energy of the system
with more than 10\% accuracy with respect to the Monte Carlo results. As such,
we expect the current theory to have a similar level of accuracy in the
unitarity limit for the $d$-wave system.

The problem that lies at the root of the difficulty to make a theory for
unitarity is that one needs to take into account not only the fluctuations of
the order parameter but also the normal-state interactions correctly. The
diagrammatic approaches \cite{strinati,levin} are based on a zeroth-order
decoupling that emphasizes pair formation rather than normal-state
interactions. Put in the language of functional integration\cite{Randeria2007}%
, we have made a particular choice for the Hubbard-Stratonovic decoupling: we
took $\bar{\psi}\bar{\psi}$ and $\psi \psi$ types of products of Grassmann
variables. This emphasizes pairing, but when the pairing goes to zero at the
saddle-point, the resulting normal state has no interactions, and fluctuation
corrections are needed to remedy this. We could have made the choice to group
$\bar{\psi}\psi$ and $\bar{\psi}\psi$ and apply the Hubbard-Stratonovic scheme
to decouple the four-product in these densities rather than in the pairs. The
resulting saddle-point approximation would yield the random phase
approximation (RPA) results for the interacting normal state; but it would
lack pairing. The inability to include --on the level of a saddle-point
approximation-- both pairing and RPA normal-state interactions through the
introduction of two collective fields is discussed by Kleinert
\cite{kleinert1}, who proposes variational perturbation theory as a
solution\cite{kleinert2}. The fluctuation expansion used in the current work
goes beyond that of Ref.\cite{Randeria2007}, in that we take not only the
particle-pair $\bar{\psi}\bar{\psi}$ and hole-pair $\psi \psi$ excitations, but
also particle-hole terms $\bar{\psi}\psi$ are present. These contributions
appear in terms that do not vanish as the saddle point goes to zero,
$\Delta \rightarrow0$, so that the normal state in the present treatment is the
interacting Fermi gas rather than the ideal Fermi gas. As such, the present
treatment will be better suited in the unitary limit.

\subsection{Routes for experimental observation}

Magnetically tuning the population imbalance in a Fermi superfluid is out of
reach at present in high-temperature superconductors. In cold atomic gases it
has been successfully demonstrated and applied to reach the superfluid regime.
The currently realized atomic Fermi superfluids have $s$-wave symmetry of the
order parameter. The $d$-wave coupling strength generally is too small to
dominate the $s$-wave scattering at low temperature. This can be overcome
using a Feshbach resonance in the $d$-wave scattering channel. $d$-wave
Feshbach resonances have been observed, for example, in various isotopes of
rubidium\cite{dwavefesh}. However, to reach the unitary limit for the $d$-wave
scattering one needs a better control over the magnetic field than in the
$s$-wave case, since the interaction parameter $\lambda,$ expression
(\ref{lambda}), scales as $a_{d}^{-5}$ as compared to $a_{s}^{-1}$. The
current results suggest a different route towards $d$-wave superfluidity:
imbalancing the gas. The dominant $s$-wave pairing is easily suppressed on the
BCS side of the resonance by adding imbalance, whereas $d$-wave superfluidity
is less sensitive to imbalance. Both the use of a $d$-wave Feshbach resonance
to obtain a large enough $d$-wave coupling strength, and of imbalance to
suppress $s$-wave pairing, will be needed to realize $d$-wave superfluidity in
the atomic gases.

In an inhomogneous trapping potential, phase separation can occur in real
space. For $s$-wave superfluidity, this leads to a balanced superfluid at the
center of the trap, surrounded by a halo of imbalanced (or fully polarized)
normal gas \cite{pilatiPRL100}. In effect, the excess spin component has been
expelled from the balanced $s$-wave superfluid. The additional energy cost of
placing the excess atoms fhigher up the trapping potential is compensated by
the energy gained by allowing the balanced superfluid state to form. In the
$d$-wave superfluid, this energy balance is different. Increasing imbalance in
the BCS side does not strongly reduce the free energy of the superfluid. As
can be seen from Fig. \ref{fig2}, the $d$-wave order parameter is not strongly
affected. Therefore, expelling the excess atoms to the edge of the trap raises
the total energy and we do not expect real-space phase separation. The
situation is different on the BEC side: here, the free energy of the $d$-wave
superfluid is reduced by imbalance, and it becomes energetically favourable to
expell excess atoms.

\subsection{Exotic pairing scenarios}

Note that the action for the fluctuations depends (through $\mathcal{M}_{11}$
and $\mathcal{M}_{21}$) on the choice of saddle point $\Delta$. This means
that the spectrum of excitations (obtained from the diagonalized fluctuation
action) also depends on the choice of the saddle point. Excitations for a
vortex condensate may be different from excitations on top of a ground-state
condensate. At nonzero temperature those excitations will be populated through
Bose statistics. But the physics is more complex than just Bose populating
excitations: the excitation spectrum itself (the dispersion and lifetime of
those excitations) is temperature dependent: new single-particle and
collective modes appear and shift as a function of temperature. At zero
temperature the only single-particle excitations are $E_{\mathbf{k}}$, the
energy spectrum for breaking a Cooper pair, but at finite temperature, we also
have the excitations of the thermal gas. These consist in taking the atoms of
a broken Cooper pair, and giving those atoms an extra kick: $E_{\mathbf{k+q}%
}-E_{\mathbf{k}}$. Besides those single-particle excitations, we will have
collective excitations whenever $\mathcal{M}_{11}\mathcal{M}_{22}%
\mathcal{-M}_{21}\mathcal{M}_{12}=0$.

In the case of an imbalanced Fermi gas, an alternative choice for the saddle
point is $\Delta e^{i\mathbf{kr}}$ where \textbf{k} represent a shortest wave
vector connecting the Fermi surface of the minority component to that of the
majority component. The resulting equations describe the
Fulde-Ferrell-Larkin-Ovchinnikov state \cite{FFLO,yoshida}. However, this
state has not yet been reported experimentally, so we have restricted the
present analysis to the usual pairing scenario.

\subsection{Conclusions}

We have investigated the imbalanced \emph{d}-wave Fermi gas, both at zero and
at nonzero temperatures, and as a function of the $d$-wave interaction
strength. We find that in the BCS regime, the $d$-wave pairing is more robust
to the presence of population imbalance than the $s$-wave case. For a range of
interaction strengths, we find that the $s$-wave superfluidity is suppressed
whereas the $d$-wave superfluidity is not. This is shown to be related to the
possibility of creating a polarized gas of excitations in the nodes of the
gap. Rather than phase separation in real space, phase separation can occur in
reciprocal space. An additional difference with the $s$-wave BCS case, is that
a critical attraction strength is needed in the $d$-wave case before pairing
can occur (in $s$-wave pairing occurs for all attractive interaction
strengths). In the BEC\ regime, the symmetry of the pairing interaction plays
a less important role: as the molecule gets more tightly bound, the details of
its internal wave function matter less, and we retrieve known results for the
$s$-wave system in the same BEC\ limit\cite{OhashiPRA67,strinati,levin}. We
then investigate how our results are affected by Gaussian fluctuations,
important both to describe the nonzero-temperature thermodynamics. Both the
critical temperature and the effect of temperature on the spectral density of
the excitations are calculated. Our investigation of the structure factor
reveals that for the $d$-wave scattering, the damping of the pole for
$S\left(  \mathbf{q},\omega \right)  $ is very small even in the BCS regime, in
contrast to that for the $s$-wave scattering. The critical temperature in the
BCS regime reflects the pair binding energy. This implies that the critical
temperature for the $d$-wave superfluid in the BCS regime will also be more
robust against population imbalance.

\begin{acknowledgments}
This work has been supported by the FWO-V Project Nos. G.0356.06, G.0115.06,
G.0435.03, G.0306.00, the WOG Project No. WO.025.99N, and the NOI BOF UA 2004.
\end{acknowledgments}

%

\appendix

\section{Matsubara summation for the density}

\begin{figure}
[h]
\begin{center}
\includegraphics[
height=2.3705in,
width=2.5425in
]%
{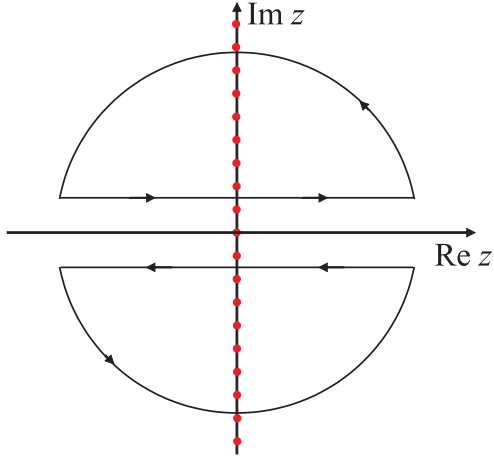}%
\caption{(color online) Integration contour in the complex $z$-plane. The full
dots indicate the poles $z=i\Omega_{n}$, $n=0,\pm1,\pm2,\ldots$.}%
\label{contour}%
\end{center}
\end{figure}

Let us consider the contour integral on the contour $C$ shown in Fig.
\ref{contour}:%
\begin{equation}
I\equiv \frac{1}{2\pi i}\oint \nolimits_{C}\frac{f\left(  z\right)  }{e^{\beta
z}-1}dz, \label{I1}%
\end{equation}
where the points $z=i\Omega_{n}$ with $\left \vert n\right \vert >n_{0}$ lie
inside the contour, and the other points $z=i\Omega_{n}$ are outside the
contour. The function $f\left(  z\right)  $ possesses the following
properties: (i) it is analytic in the entire complex $z$-plane except,
possibly, the branching line on the real axis, (ii) $f\left(  z\right)  $
decreases at $\operatorname{Re}z\rightarrow-\infty$ faster than $z^{-1}$, so
that the integral $\int_{-\infty}^{0}f\left(  \omega \pm i\gamma \right)
d\omega$, where $\omega$ and $\gamma$ are real, converges. The functions
$J\left(  \mathbf{q},z\right)  $ and $K\left(  \mathbf{q},z\right)  $
determined, respectively, by Eqs. (\ref{FJ}) and (\ref{FK}), satisfy these
conditions. The fraction $\frac{1}{e^{\beta z}-1}$ has the poles at
$z=i\Omega_{n}$, $n=0,\pm1,\pm2,\ldots$. The residues of $\frac{1}{e^{\beta
z}-1}$ in the points $z=i\Omega_{n}$ are equal to $\frac{1}{\beta}$.%

On the one hand, the integral (\ref{I1}) is equal to the sum of the residues
of the function $\frac{f\left(  z\right)  }{e^{\beta z}-1}$ in the points
$z=i\Omega_{n}$ inside the contour $C$:%
\begin{equation}
I=\frac{1}{\beta}\sum_{\left \vert n\right \vert >n_{0}}f\left(  i\Omega
_{n}\right)  , \label{q1}%
\end{equation}
On the other hand, the integral $I$ is
\begin{equation}
I=\frac{1}{2\pi i}\int_{-\infty}^{\infty}\frac{f\left(  \omega+i\gamma \right)
}{e^{\beta \left(  \omega+i\gamma \right)  }-1}d\omega-\frac{1}{2\pi i}%
\int_{-\infty}^{\infty}\frac{f\left(  \omega-i\gamma \right)  }{e^{\beta \left(
\omega-i\gamma \right)  }-1}d\omega, \label{q2}%
\end{equation}
where the parameter $\gamma$ satisfies the inequality%
\begin{equation}
\Omega_{n_{0}}<\gamma<\Omega_{n_{0}+1}. \label{ineq}%
\end{equation}
It follows from the equivalence of (\ref{q1}) and (\ref{q2}) that%
\begin{equation}
\sum_{n}f\left(  i\Omega_{n}\right)  =\frac{\beta}{\pi}\int_{-\infty}^{\infty
}\operatorname{Im}\left[  \frac{f\left(  \omega+i\gamma \right)  }%
{e^{\beta \left(  \omega+i\gamma \right)  }-1}\right]  d\omega+\sum_{n=-n_{0}%
}^{n_{0}}f\left(  i\Omega_{n}\right)  . \label{sum}%
\end{equation}

According to the theorem (\ref{sum}), the fluctuation contributions to the
density and to the population imbalance can be represented as%
\begin{align}
n_{fl}  &  =-\int \frac{d\mathbf{q}}{\left(  2\pi \right)  ^{3}}\left(  \frac
{1}{\pi}\int_{-\infty}^{\infty}\operatorname{Im}\left[  \frac{J\left(
\mathbf{q},\omega+i\gamma \right)  }{e^{\beta \left(  \omega+i\gamma \right)
}-1}\right]  d\omega+\frac{1}{\beta}\sum_{n=-n_{0}}^{n_{0}}J\left(
\mathbf{q},i\Omega_{n}\right)  \right)  ,\\
\delta n_{fl}  &  =-\int \frac{d\mathbf{q}}{\left(  2\pi \right)  ^{3}}\left(
\frac{1}{\pi}\int_{-\infty}^{\infty}\operatorname{Im}\left[  \frac{K\left(
\mathbf{q},\omega+i\gamma \right)  }{e^{\beta \left(  \omega+i\gamma \right)
}-1}\right]  d\omega+\frac{1}{\beta}\sum_{n=-n_{0}}^{n_{0}}K\left(
\mathbf{q},i\Omega_{n}\right)  \right)  .
\end{align}
As follows from the above analytical transformations of the integrals in the
complex $z$-plane, the sum (\ref{sum}) does not depend on the choice of the
number $n_{0}$ and (for a given $n_{0}$) on the value of $\gamma$ within the
range given by (\ref{ineq}).


\begin{thebibliography}{99}                                                                                               %


\bibitem[*]{jtharv}also at Lyman Laboratory of Physics, Harvard University,
Cambridge MA 02138, USA.

\bibitem {FermiBCS}C.A. Regal, M. Greiner, and D. Jin, Phys. Rev. Lett.
\textbf{92}, 040403 (2004); M. W. Zwierlein et al., Phys. Rev. Lett.
\textbf{92}, 120403 (2004); T. Bourdel et al., Phys. Rev. Lett. \textbf{93},
050401 (2004); G. B. Partridge et al., Phys.\ Rev. Lett. \textbf{95}, 020404 (2005).

\bibitem {FermiBEC}M. Greiner, C.A. Regal and D. Jin, Nature \textbf{426}, 537
(2003); S. Jochim et al., Science \textbf{302}, 2101 (2003), M.W. Zwierlein et
al., Phys.\ Rev. Lett. \textbf{91}, 120403 (2003).

\bibitem {MIT-imbal}M.W. Zwierlein, A. Schirotzek, C.H. Schunck, and W.
Ketterle, Science \textbf{311}, 492-496 (2006).

\bibitem {RICE-imbal}G.B. Partridge, W. Li, R.I. Kamar, Y.-A. Liao, and R. G.
Hulet, Science \textbf{311}, 503-505 (2006); G. B. Partridge, Wenhui Li, Y. A.
Liao, R. G. Hulet, M. Haque and H. T. C. Stoof, Phys. Rev. Lett. \textbf{97},
190407 (2006).

\bibitem {theory-imbal}A. Sedrakian \emph{et al.}, Phys. Rev. A \textbf{72},
013613 (2005); C.-H. Pao and S.-T. Wu, S.-K. Yip, Phys. Rev. B \textbf{73},
132506 (2006); D. E. Sheehy and L. Radzihovsky, Phys. Rev. Lett. \textbf{96},
060401 (2006); J. Dukelsky \emph{et al.}, ibid. \textbf{96}, 180404 (2006); P.
Pieri and G. C. Strinati, ibid. \textbf{96}, 150404 (2006); J. Kinnunen, L. M.
Jensen, and P. Torma, ibid. \textbf{96}, 110403 (2006); F. Chevy, ibid.
\textbf{96}, 130401 (2006); K. Machida, T. Mizushima, and M. Ichioka, ibid.
\textbf{97}, 120407 (2006); M. Haque and H. T. C. Stoof, Phys. Rev. A
\textbf{74}, 011602 (2006).

\bibitem {ClogstonPRL9}A.M. Clogston, Phys. Rev. Lett. \textbf{9}, 266 (1962).

\bibitem {DeSilvaPRA73}T.N. De Silva and E.J. Mueller, Phys. Rev. A
\textbf{73}, 051602 (2006).

\bibitem {FFLO}P. Fulde and R. A. Ferrell, Phys. Rev. \textbf{135}, A550
(1964); A. I. Larkin and Y. N. Ovchinnikov, Sov. Phys. JETP \textbf{20}, 762 (1965).

\bibitem {SarmaJPCS24}G. Sarma, J. Phys. Chem. Solids \textbf{24}, 1029 (1963).

\bibitem {randeria}M. Randeria, in \emph{Bose Einstein Condensation}, edited
by A. Griffin, D. Snoke, and S. Stringari (Cambridge Univ. Press, Cambridge,
1995), pp. 355--92; Q. Chen, J. Stajic, and K. Levin, Low. Temp. Phys.
\textbf{32}, 406 (2006) [Fiz. Nizk. Temp. \textbf{32}, 538 (2006)].

\bibitem {wisenature}M. C. Boyer, W. D. Wise, K. Chatterjee, M. Yi, T. Kondo,
T. Takeuchi, H. Ikuta, and E. W. Hudson, Nature Physics \textbf{3}, 802 (01
Nov 2007).

\bibitem {WilliamsPRL87}G. V. M. Williams, J. L. Tallon, E. M. Haines, R.
Michalak, and R. Dupree, Phys. Rev. Lett. \textbf{78}, 721 (1997).

\bibitem {alexandrovbook}The mounting evidence for the preformed pair scenario
in high-T$_{c}$ superconductors is described in A. S. Alexandrov, in
\emph{Theory of Superconductivity: From Weak to Strong Coupling }(IoP
Publishing, Bristol-Philadelphia, 2003), and references therein.

\bibitem {alexandrov}A. S. Alexandrov and P. E. Kornilovitch, J. Phys.
Condens.Matter \textbf{14}, 5337 (2002); A. S. Alexandrov and A. F. Andreev,
Europhys. Lett. \textbf{54}, 373 (2001).

\bibitem {ourlattice}M. Wouters, J. Tempere, J.T. Devreese, Phys. Rev. A
\textbf{70}, 013616 (2004); J. Tempere, M. Wouters and J.T. Devreese, Phys.
Rev. B \textbf{75}, 184526 (2007).

\bibitem {ourvortex}J. Tempere, M. Wouters, J.T. Devreese, Phys. Rev. A
\textbf{71}, 033631 (2005); J. Tempere, J.T. Devreese, to appear in Physica C.

\bibitem {demelo1993}C. A. R. S\'{a} de Melo, M. Randeria, and J. R.
Engelbrecht, Phys. Rev. Lett. \textbf{71}, 3202 (1993)

\bibitem {NSR}P. Nozi\`{e}res and S. Schmitt-Rink, J. Low Temp. Phys.
\textbf{59}, 195 (1985).

\bibitem {Taylor2006}E. Taylor, A. Griffin, N. Fukushima, and Y. Ohashi, Phys.
Rev. A \textbf{74}, 063626 (2006).

\bibitem {Engelbrecht1997}J. R. Engelbrecht, M. Randeria, and C. A. R. S\'{a}
de Melo, Phys. Rev. B\textbf{55}, 15153 (1997).

\bibitem {Randeria2007}R. B. Diener, R. Sensarma, and M. Randeria,
arXiv:0709.2653v1 (2007).

\bibitem {Fukushima2007}N. Fukushima, Y. Ohashi, E. Taylor, and A. Griffin,
Phys. Rev. A \textbf{75}, 033609 (2007).

\bibitem {DM2000}R. D. Duncan and C. A. R. S\'{a} de Melo, Phys. Rev. B
\textbf{62}, 9675 (2000).

\bibitem {demelopwave}M. Iskin, C.A.R. S\'{a} de Melo, Phys.\ Rev. Lett.
\textbf{96}, 040402 (2006).

\bibitem {izdiasz}Z. Izdiaszek \& T. Calarco, Phys. Rev. Lett. \textbf{96},
013206 (2006).

\bibitem {Nozieres1985}P. Nozi\`{e}res and S. Schmitt-Rink, J. Low Temp. Phys.
\textbf{59}, 195 (1985).

\bibitem {RegalPRL90}C. A. Regal, C. Ticknor, J. L. Bohn, and D. S. Jin,
Phys.\ Rev. Lett. \textbf{90}, 053201 (2003).

\bibitem {OhashiPRA67}Y. Ohashi and A. Griffin, Phys. Rev. A \textbf{67},
063612 (2003).

\bibitem {strinati}A. Perali, P. Pieri, L. Pisani, G.C. Strinati, Phys. Rev.
Lett. \textbf{92}, 220404 (2004).

\bibitem {levin}Q. Chen, Y. He, C.C. Chien, K. Levin, Phys. Rev. B
\textbf{75}, 014521 (2007).

\bibitem {MC}S. Y. Chang, V. R. Pandharipande, J. Carlson, and K. E. Schmidt,
Phys. Rev. A \textbf{70}, 043602 (2004); G. E. Astrakharchik, J. Boronat, J.
Casulleras, and S. Giorgini, Phys. Rev. Lett. \textbf{93}, 200404 (2004).

\bibitem {kleinert1}H. Kleinert, Fortschr. Physik \textbf{26}, 55 (1978).

\bibitem {kleinert2}H. Kleinert, Annals of Physics \textbf{266}, 135 (1998).

\bibitem {pilatiPRL100}S. Pilati and S. Giorgini, Phys. Rev. Lett.
\textbf{100}, 030401 (2008).

\bibitem {dwavefesh}H. M. J. M. Boesten et al., Phys. Rev. A \textbf{55}, 636
(1997); J. P. Burke and J. L. Bohn, Phys. Rev. A \textbf{59}, 1303 (1999).

\bibitem {yoshida}N. Yoshida and S.-K. Yip, Phys. Rev. A \textbf{75}, 063601 (2007)
\end{thebibliography}
\end{document}